\documentclass[12pt]{amsart}
\usepackage{amssymb}
\usepackage{latexsym}
\usepackage{epic, eepic, graphicx}
\usepackage{graphicx} \DeclareGraphicsExtensions{.ps}

\newcommand{\nwc}{\newcommand}

\nwc{\IA}{\mathbb{A}} 
\nwc{\IB}{\mathbb{B}} 
\nwc{\IC}{\mathbb{C}} 
\nwc{\ID}{\mathbb{D}} 
\nwc{\IE}{\mathbb{E}} 
\nwc{\IF}{\mathbb{F}} 
\nwc{\IG}{\mathbb{G}} 
\nwc{\IH}{\mathbb{H}} 
\nwc{\IN}{\mathbb{N}} 
\nwc{\IP}{\mathbb{P}} 
\nwc{\IQ}{\mathbb{Q}} 
\nwc{\IR}{\mathbb{R}} 
\nwc{\IS}{\mathbb{S}} 
\nwc{\IT}{\mathbb{T}} 
\nwc{\IZ}{\mathbb{Z}} 

\nwc{\cO}{\mathcal{O}}

\nwc{\eps}{\epsilon}
\nwc{\cP}{\mathcal{P}}

\nwc{\tr}{\operatorname{tr}}
\nwc{\Res}{\operatorname{Res}}
\nwc{\Spec}{\operatorname{Spec}}
\nwc{\Vol}{\operatorname{Vol}}
\nwc{\Var}{\operatorname{Var}}

\nwc{\la}{\langle}
\nwc{\ra}{\rangle}

\renewcommand{\Re}{\operatorname{Re}}
\renewcommand{\Im}{\operatorname{Im}}

\title{Some open questions in ``wave chaos"}
\author{St\'ephane Nonnenmacher}
\address{Institut de Physique Th\'eorique, 
CEA, IPhT, F-91191 Gif-sur-Yvette, France;
CNRS, URA 2306, F-91191 Gif-sur-Yvette, France}

\begin{document}


\maketitle

The subject area referred to as ``wave chaos", ``quantum chaos" or ``quantum chaology" has been investigated mostly by the theoretical physics community in the last 30 years. The questions it raises have more recently also attracted the attention of mathematicians and mathematical physicists, due to connections with number theory, graph theory, Riemannian, hyperbolic or complex geometry, classical dynamical systems, probability etc. After giving a rough account on ``what is quantum chaos?", I intend to list some pending questions, some of them having been raised a long time ago, some others more recent. 
The choice of problems (and of references) is of course partial and personal.\footnote{This contribution was mostly written while I enjoyed a stay at the Centre de Recherches Math\'ematiques in Montr\'eal, invited by D.Jakobson and I.Polterovich, to whom I am thankful. My recent research was supported by the grant ANR-05-JCJC-0107-01.}

\subsection{A brief overview of ``quantum chaos"}
Let us start by resolving the apparent paradox of the phrase ``quantum chaos".
A classically chaotic system is defined by an ordinary differential equation, or a discrete-time map acting on some finite-dimensional phase space; for the dynamics to be chaotic, the ODE or the map is generally {\em nonlinear} in the phase space coordinates. By contrast, the associated quantum system is defined by a {\em linear} operator acting on the quantum Hilbert space; the quantum dynamics generated by this operator (either a selfadjoint Hamiltonian, or a unitary propagator) is by essence linear, but acts in a space of infinite (or at least large) dimension. Let us assume that the spectrum of that operator is discrete: the quantum dynamics is then quasiperiodic, and can be easily expressed in terms of the eigenvalues and eigenstates of the operator. One is thus led, from a time-dependent (dynamical) problem, to a time-independent (spectral) problem, which forms the ``backbone" of quantum chaos:
\begin{quote}
What are the spectral properties of a quantum Hamiltonian, 
the classical limit of which generates a chaotic dynamics? \\

What do the corresponding eigenstates look like?
\end{quote}
These two questions are deliberately vague. Below I will give more precise formulations. 
To fix ideas, we will consider one of the simplest chaotic systems, namely the geodesic flow on a compact Riemannian surface $(X,g)$ of {\em negative curvature} (and {\em unit area}). The corresponding quantum Hamiltonian is defined as the generator of the wave group on $X$, namely $P=\sqrt{-\Delta}$, where $\Delta$ is the Laplace-Beltrami operator. $P$ has discrete spectrum $\{\lambda_k,k\in\IN\}$ associated with eigenfunctions $\{\psi_k\}$.

The first question one can ask is ``Do we have exact, or at least asymptotic formulas to describe the high-energy eigenvalues/eigenfunctions?" For chaotic systems\footnote{Some particular quantum chaotic maps, like the quantum cat map, admit explicit expressions for eigenvalues and eigenstates. However, this system is considered to be  ``nongeneric" in most respects.}, and as opposed to the case of completely integrable systems (for which the Maslov-WKB method provides asymptotic expansions to any order in $\lambda^{-1}$), the answer is ``No".

\section{Spectral statistics}
We first investiate the spectrum $\{\lambda_k\}$ of the quantum Hamiltonian $P$.
Although there are no asymptotic expressions for the eigenvalues,
one can still obtain some crude information on their distribution, especially in the high-energy limit $\lambda_k\to\infty$. Indeed, this limit is a {\em semiclassical} limit, where the classical dynamics is expected to be relevant.

\subsection{Macroscopic spectral properties}
Firstly, one can investigate the ``macroscopic" spectral distribution. For any surface of unit area, the behaviour of the spectral counting function
$N(\lambda)=\#\{k\in\IN,\ \lambda_k\leq \lambda\}$ is asymptotically given by the celebrated Weyl's law
\begin{equation}\label{Weyl}
N(\lambda)=\frac{\lambda^2}{4\pi}+R(\lambda),\ \text{with a remainder}\quad R(\lambda)=\cO(\lambda)\ \text{when}\ \lambda\to\infty.
\end{equation}
The size of the remainder gives an idea of the ``clustering" of the eigenvalues. Maximal clustering appears e.g. for the standard sphere, where\footnote{We recall that $R(\lambda)=\Omega(f(\lambda))$ means that $\limsup_{\lambda\to\infty}\frac{|R(\lambda)|}{f(\lambda)}>0$.} $R(\lambda)=\Omega(\lambda)$. 
The dynamics of the geodesic flow is ``invisible" in the main term, but already pops up in the remainder term: for a negatively curved surface, one can prove that $R(\lambda)=\cO(\lambda/\log\lambda)$ \cite{Berard77}; one actually expect that, for a ``generic" such surface, $R(\lambda)=\cO(\lambda^\eps)$ for any $\eps>0$. 

On the other hand, for {\em arithmetic} surfaces, it is known that $R(\lambda)=\Omega(\sqrt{\lambda}/\log\lambda)$, indicating a higher amount of clustering.
In varying negative curvature, the smaller lower bound $R(\lambda)=\Omega((\log\lambda)^{\cP(-1/2\mathcal{H})/\cP(0)-\eps})$ was recently shown in \cite{JPT}. Here the (positive) exponent is given in terms of the topological pressure $\cP(\bullet)$, a quantity characterizing the classical dynamics. 

To prove such spectral estimates, one applies a semiclassical approximation for the trace of the propagator $\tr (e^{itP})$, in the form of a sum over closed geodesics (the Gutzwiller trace formula). A Fourier transform then maps the time-dependent traces to energy-dependent ones, to finally yield $N(\lambda)$. 
The rule of the game is: the further in time you control the propagator, the more precise your spectral estimate becomes. For a chaotic system, trajectories separate from each other at an exponential rate, which results in a breakdown of semiclassical approximations around the {\em Ehrenfest time} $T_E\approx \frac{\log\lambda}{\chi_{Lya}}$, where $\chi_{Lya}$ is the largest Lyapunov exponent. In that view, a present line of investigation consists in trying to extend the validity of the Gutzwiller approximation to larger times \cite{Faure07}.

The result of \cite{JPT} is characteristic of ``quantum chaos": to obtain spectral estimates which are sharper than for general surfaces, one needs to take into account some refined properties of the classical dynamics. The topological pressure, the entropy, the Ruelle-Pollicott spectrum associated with the geodesic flow, should ``naturally" arise in the description of the quantum spectrum.

\subsection{Microscopic spectral statistics}
The Bohigas-Giannoni-Schmit (BGS, or ``Random matrix theory", RMT) conjecture \cite{BGS84} aims at describing the spectral distribution in a finite window around $\lambda\gg 1$ at the {\em microscopic} scale, namely the scale of the mean level spacing $\delta\lambda\sim 2\pi/\lambda$. At this scale, the spectrum resembles an infinite sequence of ``pseudorandom points". The BGS conjecture claims that, provided the classical dynamics is chaotic, this sequence is statistically equivalent with that formed by the eigenvalues of a random matrix belonging to one of the three standard Gaussian ensembles (in the present case the Gaussian Orthogonal Ensemble, made of real-symmetric matrices with independent Gaussian entries). Thus, the local spectral statistics is ``universal".
The statistical observables studied to test this conjecture are, e.g., the 2-point correlation function and the level-spacing distribution, which are known analytically in the case of the random matrix ensembles \cite{Mehta}.

Attempts have been made to semiclassically compute the 2-point correlation function, or equivalently its Fourier transform called the spectral form factor $K(t)\propto \la |\tr (e^{itP})|^2 \ra$. The most popular method (initiated by Berry) uses Gutzwiller's approximate expression for the trace: taking its square results in a sum over {\it pairs} of periodic orbits
$K(t)\propto \sum_{\gamma,\gamma'}A_\gamma \bar{A}_{\gamma'} e^{i\lambda(\ell_\gamma-\ell_{\gamma'})}$ with lengths $\ell_\gamma+\ell_{\gamma'}$ close to the {\em Heisenberg time} $T_H \sim \lambda$. Berry assumed that, upon some averaging over $\lambda$, only the pairs of equal lengths $\ell_\gamma=\ell_{\gamma'}$ contribute to the sum \cite{Berry85}. More recently, Sieber and Richter identified pairs of almost identical lengths \cite{SR01}, which also contribute. Elaborating on their insight (which implies some nontrivial combinatorics), the Essen group (+ guests) managed to reproduce the RMT form factor from the periodic orbit expansion \cite{HMABH07}. 

Yet, this remarkable result is still far from achieving a rigorous proof of the BGS conjecture. Firstly, the Gutzwiller approximation has been applied to times of order $T_H\gg T_E$, for which its validity is only speculative (except for some special systems, like surfaces of constant curvature, the quantum cat map, or quantum graphs). The second problem seems more formidable: one needs to prove that the huge sum of ``non-correlated pairs" is effectively negligible compared to that over correlated ones. Considering our knowledge of the orbit length distribution, 
this objective seems presently out of reach \cite{PollSharp06}.

In the case of arithmetic surfaces (or the quantum cat map), the orbit lengths are highly degenerate (they are simple functions of integer numbers), so many more pairs of orbits are correlated than in the generic case. The above approach implies that the 2-point function converges to that of Poisson-distributed points, quite different from RMT \cite{BGGS92}; this untypical statistics has been checked numerically, but a full proof of is still lacking
(mainly due to the ``second problem" mentioned above) \cite{Peter02}.

\subsection{"Generic" vs. ``arithmetic"}
\subsubsection{Generic surfaces} Among the surfaces of negative curvature, some possess ``arithmetic symmetries" in the form of a commutative algebra of ``Hecke correspondences". These symmetries are not easily visible when characterizing the ergodic properties of the classical system (like exponential mixing); they have more subtle consequences, like large degeneracies of the ``length spectrum" $\{\ell_\gamma\}$, which, as we have seen, strongly impact the quantum spectrum at the microscopic scale. Although the geodesic flow is fully chaotic, the quantum spectrum is ``nonuniversal". 

The existence of such ``exceptional" surfaces invalidates any proof of the BGS conjecture which would be only based on global chaotic properties. To prove the conjecture, one will need to exhibit a more refined dynamical property able to explicitly separate the exceptional systems from the generic ones. The above results suggest that this property may concern the length spectrum: it should probably forbid large degeneracies length degeneracies, but must also deal with almost-degeneracies, maybe  by requiring some diophantine condition on the differences $\{\ell_\gamma-\ell_{\gamma'}\}$.

Once such a property has been identified, one could ask: 
\begin{quote}
Among all surfaces of negative curvature, how generic are those satisfying this property?
\end{quote}
In the end, one can hope to prove the BGS conjecture for a ``generic" negatively curved surface, where ``generic" could be understood in a topological, or a measure-theoretical sense. Let us mention that ``generic" properties of classical dynamical systems have been investigated for years (see e.g. \cite{BoVia04}).

\subsubsection{Quantum genericity}
So far the quantum Hamiltonian associated with the geodesic flow on $(X,g)$ has always been the Laplace-Beltrami operator. Yet, a perturbation of the Laplacian  by (selfadjoint) lower-order terms still makes up a ``quantization" of the geodesic flow. More generally, the ``quantization" of a Hamiltonian dynamics proceeds by {\em choosing} a certain quantization scheme, different schemes leading to different operators with different spectra. The problem of ``genericity" then may as well concern the various quantizations attached to a given classical dynamics \cite{Wilkin88}. 
For instance, if one perturbs the Laplacian on an arithmetic surface by ``generic" lower-order terms, does the resulting spectrum become universal? If yes, how large must be the perturbing terms to achieve ``universality"?

\section{Anatomy of chaotic eigenstates}

Let us now consider a different question, namely the description of the (high-energy) eigenfunctions $\psi_k(x)$ of the quantum system (we still stick to the Laplacian on $(X,g)$). There are many ways to analyze those functions, we will mention only some of them.

\subsection{Macroscopic distribution of the eigenmodes: quantum ergodicity}

Since $|\psi_k(x)|^2$ represents the probability density of the particle to sit at the position $x$, one first wonders whether a ``stationary quantum particle" can preferably be localized in some region $A\subset X$, in the sense that\footnote{We recall that the area of $X$ is unity, and the eigenstates $\psi_k$ are $L^2$-normalized.} $\int_{X\setminus A} |\psi_k(x)|^2dx \ll  \Vol(X\setminus A)$. This seems counter-intuitive, considering the ergodicity of the classical flow. Shnirelman indeed showed that {\it almost all} high-energy eigenstates are asymptotically equidistributed, in the sense that 
$\int_{A} |\psi_k(x)|^2dx \approx \Vol(A)$ \cite{Shnir}. 

To make the connection with classical dynamics, it is convenient to lift the wavefunction $\psi_k(x)$ into the phase space $T^*X$ \cite{Berry77,Voros77}. There exist various ways to associate with $\psi_k$ a  ``semiclassical measure" $\mu_k$, which also describes the way the wavefunction is distributed in ``momentum at the scale $\lambda_k$".
The extension of Shnirelman's result to a phase space setting reads as follows: there exists a subsequence $S\subset\IN$ of full density, such that the measures $(\mu_k)_{k\in S}$  weak-$*$ converge to $\mu_{L}$, the Liouville measure on the unit cotangent bundle $S^*X$ \cite{CdV85}.
This property of ``generic high-energy eigenstates" is called Quantum Ergodicity. Its proof is robust: it only uses classical ergodicity of the dynamics, and some basic properties of quantization.

\subsubsection{Distribution of quantum averages}
At which {\it rate} does this convergence take place?
Namely, for a given phase space observable $f$, how fast does $\mu_k(f)\to\mu_{L}(f)$? One would actually like to understand the {\it distribution} of the elements $\{\mu_k(f),\ \lambda_k\leq \lambda\}$ in the limit $\lambda\to\infty$. The proof of quantum ergodicity proceeds by showing that the variance of the distribution (sometimes called ``Quantum variance") vanishes when $\lambda\to\infty$. If the classical flow is {\it fast enough mixing}, the best proven upper bound for the variance is $\cO(1/\log\lambda)$ \cite{Zel94}. On the other hand, based of Random Matrix arguments and some form of Central Limit Theorem, Feingold and Peres have conjectured that, for a generic chaotic system, the distribution should be Gaussian with  variance $\sim \Var(f)/\lambda$, where $\Var(f)$ is the variance appearing in the {\em classical} Central Limit Theorem \cite{FP86}. This conjecture has been numerically checked for numerous nonarithmetic systems \cite{EFKAMM95}. The gap between the expected $\cO(\lambda^{-1})$ and the proved $\cO(1/\log\lambda)$ is (once more) due to our present inability to control the quantum evolution beyond the Ehrenfest time $T_E\sim \log\lambda$.

The only systems for which the quantum variance could be asymptotically computed are arithmetic: Luo and Sarnak showed that the quantum variance for the modular surface (for a Hecke eigenbasis) behaves as $V(f)/\lambda$, but with a factor $V(f)$ slightly different from $\Var(f)$. Kurlberg and Rudnick obtained the same type of estimate for the quantum variance of arithmetic (Hecke) eigenbases of the quantum cat map, for which there exist explicit formulas \cite{KR05}; they also show that the distribution is NOT Gaussian. This exemplifies a recurrent ``dilemma" of quantum chaos: the systems for which some computations can be performed analytically appear to be systematically ``nongeneric".

\subsubsection{Quantum Unique Ergodicity vs. exceptional eigenstates}
On the other hand, Schubert 
showed that, for some (nonarithmetic) eigenbases of the same quantum cat map, the quantum variance can behave as ``badly" as $C/\log\lambda$ \cite{Schu05}. In that case, the large fluctuations of the $\{\mu_k(f)\}$ seem due to the possibility, for this very special system, to construct  sequences of eigenstates which are NOT equidistributed, but for which $\mu_k\to \frac{1}{2}(\mu_{L}+\nu)$, where $\nu$ can be any classically invariant probability measure \cite{FNDB}.
Such states belong to the ``density zero" subsequence of ``exceptional eigenstates" allowed by Shnirelman's theorem. 
So far, such exceptional sequences could be constructed only for some very special quantum maps or quantum graphs.

Rudnick and Sarnak have conjectured that exceptional sequences cannot exist on negatively curved manifolds \cite{RudSar94} (that is, the full sequence $(\mu_k)_{k\in\IN}$ converges to $\mu_L$), this property being called ``Quantum Unique Ergodicity" (QUE). This non-existence was proved recently by Lindenstrauss for the Hecke eigenstates of arithmetic surfaces \cite{Linden06}\footnote{To be more precise, the surface must be obtained by quotienting the Poincar\'e half-plane by a {\it congruent} subgroup of $SL_2(\IR)$.}. Lindenstrauss's result has been extended to some higher-dimensional arithmetic manifolds \cite{SilVen07}. His proof uses the fact that the $\psi_k$ are eigenstates of an infinite set of Hecke operators, yet it is somewhat expected that a {\em single} Hecke operator (that is, a single ``arithmetic symmetry") should suffice to apply the same strategy of proof\footnote{L. Silberman, private communication.}.

The QUE conjecture is still wide open for non-arithmetic surfaces (and higher-dimensional manifolds). Rather than proving that all semiclassical measures converge to $\mu_{L}$, one can at least try to obtain some information on the possible limit measures. Such measures are invariant w.r. to the geodesic flow, which still leave a large choice. Recently, some constraints were obtained in the form of non-trivial lower bounds for the Kolmogorov-Sinai entropy of the limit measures \cite{AN07}: in the case of constant negative curvature, the entropy must be at least {\em half} its maximal value, showing that the eigenstates are ``at least half-delocalized" (to prove
QUE along these lines one would need to remove the {\em half} from the lower bound). 

On the other hand, disproving QUE amounts to exhibiting a negatively curved manifold and a sequence of exceptional eigenstates of the Laplacian, or of some perturbation of it by lower-order terms. Like the BGS conjecture, it is possible that QUE holds only for ``generic" negatively curved surfaces, or ``generic" perturbations of the Laplacian.

Finally, one may wonder how chaotic a system should be to satisfy QUE.
Does it need to be uniformly hyperbolic (Anosov), like geodesic flows in negative curvature? Would it be easier to construct exceptional states in case the manifold has some ``flat parts"? The latter question is related with the possible existence of {\it bouncing ball modes} in the stadium billiard (which have been clearly identified in numerics) or on similar manifolds \cite{Donn03}.

\subsection{Beyond the macroscopic features: chaotic=random?}

Even if true, QUE provides a rather poor description of the eigenfunctions (to be compared, for instance, with the precision of WKB approximations). One would also wish to analyze the wavefunctions $\psi_k(x)$ at {\em microscopic} scales, or at least some intermediate scale between $1$ and the wavelength $\lambda^{-1}$. 
The main conjecture concerning these microscopic properties is analogous to the BGS conjecture: the eigenfunctions are expected to be statistically equivalent with ``random states" in some appropriate ensemble. This conjecture was first formulated by Berry \cite{Berry77}, who compared the 
$\psi_k$ with random superpositions of isoenergetic plane waves, and analyzed the short-distance correlation functions of the latter. This conjecture has been extended to  quantum maps (the ensembles of random states are somewhat easier to describe in this framework).

The statistical observables to be compared with the random models are manifold. They include the value distribution for $\{\psi_k(x),\, x\in X\}$, correlation functions at distances $\sim \lambda^{-1}$, the structure of the {\em nodal set} $\{x\,|\,\psi_k(x)=0\}$ and of the complementary nodal components. One can analyze eigenstates individually, or by considering a large bunch of them in some energy window. 

To my knowledge, this field of research has mostly consisted in analyzing the properties of the random models, and then numerically checking that these properties are (approximately) satisfied by the eigenstates of some selected quantized chaotic systems. Certain global quantities, like the {\em moments} of the distribution $\{\psi_k(x),\, x\in X\}$ (that is the $L^p$ norms 
$\|\psi_k\|_p$) have also been investigated analytically. As was the case for the remainder $R(\lambda)$ of the Weyl formula \eqref{Weyl}, one can obtained
upper bounds valid for any Riemannian surface \cite{Sog88}, which are usually improved by some negative power of $\log\lambda$ in the negative curvature case. Still, these upper bounds are expected to be much larger than the $L^p$ norms of ``random states", which are expected to be valid for ``generic" negatively curved surfaces.
One the other hand sharper upper and lower bounds were obtained for the norms of Hecke eigenfunctions on some arithmetic surfaces or manifolds \cite{IwaSar95}; in dimension $\geq 3$, one can exhibit such Hecke eigenstates with ``exceptionally" large sup-norms \cite{RudSar94}. The value of distribution of Hecke eigenstates of quantum cat maps is much better understood: their $L^\infty$ norms, for instance, satisfy sharp upper bounds \cite{KR01}.

\section{"Open" quantum chaos}

Quantum chaos is also concerned with the properties of certain types of nonselfadjoint operators. One instance originates from the study of quantum scattering, in situations where the corresponding classical scattering potential leads to some chaotic dynamics. The basic example of such systems is the scattering by 3 or more disks on the plane. Provided none of the disks ``shadows" the other ones, the scattering trajectories are organized around a fractal {\em trapped set} $K\subset S^*X$, consisting of points remaining in the (compact) ``interaction region" at all times. At the quantum level, the square-root $P$ of the (Dirichlet) Laplacian outside the obstacles has a purely continuous spectrum on the real axis, but its resolvent $(P-\lambda)^{-1}$ admits a meromorphic continuation from $\{\Im\lambda>0\}$ across the real axis into the ``unphysical sheet"; each pole $\lambda_k$ of the continued resolvent is a {\em resonance}, associated with a {\em metastable state} $\psi_k$, which decays with a half-time $\propto 1/|\Im\lambda_k|$.

Through a complex scaling procedure, one can deform $P$ into a nonselfadjoint operator $P_\theta$, which admits the $\{\lambda_k\}$ as {\em bona fide} eigenvalues, associated with $L^2$ eigenstates $\tilde\psi_k$ obtained by deforming the $\psi_k$. 
One can now ask similar questions as in the selfadjoint case: 
\begin{quote}
What is the connection between the distribution of high-energy resonances and the eigenstates $\tilde\psi_k$ on one side, the chaotic dynamics on the trapped set $K$ on the other side? 
\end{quote}

\subsection{Fractal Weyl laws}
These questions have received less attention than their ``selfadjoint" counterparts, and even conjectures are sparse. An interesting aspect concerns the number of resonances in the strip $N(\lambda,C)=\#\{\lambda_k,\ |\Re\lambda_k|\leq \lambda, \Im\lambda_k\geq -C\}$, which has been first investiated by Sj\"ostrand \cite{SjDuke}. One expects a fractal Weyl law 
\begin{equation}\label{Weyl2}
N(\lambda,C)\sim f(C)\,\lambda^\nu,\quad\lambda\to\infty,
\end{equation}
where the fractal exponent $\nu=\frac{1+dim(K)}2$ is related with the (Minkowski) dimension of the trapped set, and $f(C)$ is some function of the strip width. This estimate interpolates between the case of a compact surface ($dim(K)=3$, see Eq. \eqref{Weyl}) and that of a trapped set consisting in a single hyperbolic geodesic ($dim(K)=1$), for which the resonances form a deformed lattice. 
Because the metastable states are ``living" on $K$ (see below), the above estimate naively originates from counting how many states (each occupying a phase space region of size $\lambda^{-2}$) can be squeezed in the $\lambda^{-1/2}$-neighbourhood of $K$. Yet, the spectra of nonselfadjoint operators are notoriously more difficult to analyze than their selfadjoint counterparts, which explains why only the upper bound in \eqref{Weyl2} could be proved \cite{SjDuke}. 

An Ansatz for the ``shape function" $f(C)$ in \eqref{Weyl2} was proposed by Schomerus and Tworzyd{\l}o, in terms of an ensemble of (subunitary) random matrices \cite{SchTw}; so far, the numerical data performed on quantum maps are not very conclusive. Even numerically checking the exponent $\nu$ is a nontrivial task when one deals with a real scattering system \cite{Lin02}.

The only system for an asymptotic of the form of \eqref{Weyl2} could be proved is a simplistic ``open quantum baker's map" \cite{NZ1}. Even in that model, the resonance spectrum may, for some choices of parameters, degenerate to a much smaller density (equivalent with the case of a single periodic geodesic). This shows that the estimate \eqref{Weyl2} is probably only true for ``generic" scattering systems, or ``generic quantizations" of them.

Considering the difficulty of settling the validity of this ``Weyl law", analyzing the finer (e.g. ``microscopic") distribution properties of the resonance spectrum seems presently quite difficult.

\subsection{Structure of metastable states}

The metastable states $\psi_k$ associated with resonances in the above strip can also be described through semiclassical measures $\mu_k$. Since $\psi_k$ is not normalizable, the measure $\mu_k$ is infinite; yet, only the part lying in the ``interaction region" is physically interesting (one can alternatively consider the measure $\tilde\mu_k$ associated with the deformed $L^2$ eigenstates $\tilde\psi_k$). 
The structure of $\mu_k$ in that region is intimately connected with the classical trapped set $K$: any limit measure $\mu$ of some subsequence $(\mu_{k_j})$ is supported on the unstable manifold of the trapped set; $\mu$ is not invariant through the geodesic flow, but decays at the rate $\Lambda=-2\lim_k \Im\lambda_{k_j}$.
 As opposed to the case of compact manifolds (where the Liouville measure prevails as a ``limit semiclassical measure"), in the present case quantum mechanics does not seem to ``preferably select" any limit measure. 
Some numerical (and analytical) results in the framework of ``open quantum maps" seem to rather indicate a ``democracy", with a multiplicity of semiclassical phase space distributions \cite{KNPS06}, but the general situation is still unclear at the moment.

\end{document}